\documentstyle[12pt]{article}
\def \bep{\mbox{\boldmath $\epsilon$}}
\def \beq{\begin{equation}}

\def \dg{\Delta \Gamma}
\def \eeq{\end{equation}}
\def \gb{\Gamma(B^0)}
\def \gdc{\Gamma(D^+)}
\def \gdn{\Gamma(D^0)}
\def \glb{\Gamma(\Lambda_b)}
\def \glc{\Gamma(\Lambda_c)}
\def \lb{\Lambda_b}
\def \lc{\Lambda_c}

\def \psbu{|\Psi(0)|^2_{bu}}
\def \psbub{|\Psi(0)|^2_{b \bar u}}

\def \pscdb{|\Psi(0)|^2_{c \bar d}}

\def \sb{\Sigma_b}
\def \sbs{\Sigma_b^*}
\def \sc{\Sigma_c}
\def \scs{\Sigma_c^*}

\def \tb{\tau(B^0)}

\def \tlb{\tau(\Lambda_b)}

\begin{document}
\topskip 2cm 
\begin{titlepage}
\vspace{-3in}
\rightline{DESY-96-062}
\rightline{CERN-TH/96-100}
\rightline{EFI-96-11}
\rightline{hep-ph/9604319}
\rightline{April 1996}
\begin{center}
{\large\bf TWO TOPICS IN $b$ PHYSICS\footnote{Presented at 10th Rencontre
de La Thuile, Val d'Aosta, Italy, March 4-9, 1996.}} \\
\vspace{1.5cm}
{\large Jonathan L. Rosner} \\
\vspace{.5cm}
{\sl Deutsches Elektronenen Synchrotron DESY, D-22603 Hamburg, Germany
\footnote{Address during preparation of this report.}} \\
\vspace{.5cm}
{\sl CERN, 1211-CH Geneva 23, Switzerland
\footnote{Address during presentation.}} \\
\vspace{.5cm}
{\sl Enrico Fermi Institute and Department of Physics}\\
{\sl University of Chicago, Chicago, IL 60637 USA
\footnote{Permanent address.}}\\
\vspace{2.5cm}
\vfil
\begin{abstract}

(1) A simple transversity analysis permits one to separate the P-even and P-odd
partial waves in such decays as $B_s \to J/\psi \phi$ and $B \to J/\psi K^*$. 
This method is relevant to the separation of contributions of CP-even and
CP-odd final states in $B_s$ decays, and hence to the measurement of a possible
lifetime difference between mass eigenstates. (2) The enhancement $\Delta
\Gamma (\Lambda_b)$ of the $\Lambda_b$ decay rate due to four-fermion processes
is calculated in terms of the $\Sigma_b^* - \Sigma_b$ hyperfine splitting, the
$B^* - B$ hyperfine splitting, and the $B$ meson decay constant $f_B$.  Despite
a relatively large hyperfine splitting observed by the DELPHI Collaboration,
the mechanism falls far short of being able to explain the observed
enhancement. 

\end{abstract}

\end{center}
\end{titlepage}

\section{Introduction}

Recent experimental results on $b$ physics include the first report of an
angular distribution analysis for the decay $B_s \to J/\psi \phi \to e^+ e^-
K^+ K^-$ \cite{CDFdist}, and a persistent difference between the $\Lambda_b$
lifetime (around 1.2 ps) and those of the $B$ mesons (at least 1.5 ps)
\cite{Sharma}.

In collaboration with A. Dighe, I. Dunietz, and H. Lipkin \cite{DDLR}, we have
used a transversity analysis \cite{Tr} to separate out partial waves of even
and odd parity in decays of pseudoscalar mesons to pairs of vector mesons. This
analysis, described in Section 2, permits separation of CP-even and CP-odd mass
eigenstates of the $B_s$ meson in the decay $B_s \to J/\psi \phi$. By applying
a similar analysis to obtain the relative contributions of $S + D$- vs.
$P$-wave decays in $B \to J/\psi K^*$ (for which the present data sample is
much more copious), and using flavor SU(3), one can obtain related information.

Some time ago it was suggested \cite{BLS} that the enhanced decay rate
of $\Lambda_c$ was due to the weak scattering process $c d \to s u$.  We find
\cite{JRLB} that the corresponding effect in the $\Lambda_b$ falls short of
explaining the observed rate enhancement. A similar conclusion has been reached
by others \cite{NSNGU}. As described in Sec.~3, we estimate the four-quark
matrix element using the hyperfine splitting between the $\Sigma_b^*$ and the
$\Sigma_b$ reported by the DELPHI Collaboration \cite{DELPHI}.  Although one
thus obtains a value of the four-quark matrix element which is at least twice
as large as a na\"{\i}ve estimate, it falls far short of what is needed. 

We summarize in Section 4.

\section{Angular distributions and lifetime differences}

The strange $B$ meson $B_s \equiv \bar b s$ and its charge-conjugate $\bar B_s
\equiv b \bar s$ are expected to mix with one another with a large amplitude.
The mass eigenstates $B_s^H$ (``heavy'') and $B_s^L$ (``light'') with masses
$m_H$ and $m_L$ are expected to be split by $\Delta m \equiv m_H - m_L \approx
25 \bar \Gamma$, give or take a factor of two \cite{JRCP}, where $\bar \Gamma
\equiv (\Gamma_H + \Gamma_L)/2 \approx \Gamma(B^0)$ $(B^0 \equiv \bar b d)$.
The measurement of such a large mass difference poses an experimental
challenge. 

Aside from small CP-violating effects, the mass eigenstates correspond to those
$B_s^{(\pm)}$ of even and odd CP. The decay of a $\bar B_s$ meson via the quark
subprocess $b (\bar s) \to c \bar c s (\bar s)$ gives rise to predominantly
CP-even final states \cite{CPeven}. Thus the CP-even eigenstate should have a
greater decay rate. An explicit calculation \cite{Blifes} gives 
\beq \label{eqn:widthdiff}
\frac{\Gamma(B_s^{(+)}) - \Gamma(B_s^{(-)}) }{\overline \Gamma} \simeq 0.18
\frac{f_{B_s}^2}{(200~\mathrm{MeV})^2}~~~,
\eeq
where $f_{B_s} \approx 200$ MeV is the $B_s$ decay constant (in a normalization
in which $f_\pi = 132$ MeV).  The estimate of $f_{B_s}$ \cite{JRCP,lat} is
probably good to about 20\%.

The ratio of the mass splitting to the width difference of strange $B$'s is
predicted to be large and independent of CKM matrix elements \cite{IsiBs,BP}:
$\Delta m/\Delta \Gamma \simeq {\cal O}(-[1/\pi] [m_t^2/m_b^2]) \simeq - 200$,
where $\Delta \Gamma \equiv \Gamma_H - \Gamma_L$. In view of the sign in
Eq.~(\ref{eqn:widthdiff}) and since $\Delta m >0$ by definition, we then
identify $B_s^L = B_s^{(+)}$ and $B_s^H = B_s^{(-)}$ \cite{IsiBs}. If the mass
difference $\Delta m$ turns out to be too large to measure at present because
of the rapid frequency of $B_s - \bar B_s$ oscillations it entails, the width
difference $\Delta \Gamma$ may be large enough to detect
\cite{CPeven,Blifes,largedg}. 

One can measure $\bar \Gamma$ using semileptonic decays, while decays to CP
eigenstates can be measured by studying the correlations between the
polarization states of the vector mesons in $B_s^{(\pm)} \to J/\psi \phi$. In
this section we describe a means by which the $J/\psi \phi$ final states of
definite CP in $B_s$ decays may be separated from one another using an angular
distribution based on a {\it transversity} variable \cite{Tr}.  This variable
allows one to directly separate the summed contribution of the even partial
waves (S,~D) from the odd one (P) by means of their opposite parities. 

Consider the final state $J/\psi \phi \to \ell^+ \ell^- K^+ K^-$, where $\ell =
e$ or $\mu$. In the rest frame of the $J/\psi$ let the direction of the $\phi$
define the $x$ axis. Let the plane of the $K^+ K^-$ system define the $y$ axis,
with $p_y(K^+) > 0$, so the normal to that plane defines the $z$ axis.  (We
assume a right-handed coordinate system.) We define the angle $\theta$ as the
angle between the $\ell^+$ and the $z$ axis.  Then the time-dependent rate for
the $J/\psi \phi$ mode is given by 
\beq \label{eqn:dist}
\frac{d^2 \Gamma}{d \cos \theta~dt} = \frac{3}{8} p(t) \left( 1 + \cos^2 \theta
\right) + \frac{3}{4} m(t) \sin^2 \theta~~~,
\eeq
where
\beq
p(t) = p(0) e^{-\Gamma_L t}~~~\mathrm{(CP~even)}~~~,~~~
m(t) = m(0) e^{-\Gamma_H t}~~~\mathrm{(CP~odd)}~~~,
\eeq
so that the probability of having a CP-even [CP-odd] state at proper time $t$
is given by $p(t)/(p(t) + m(t))$ [$m(t)/(p(t) + m(t))$]. The angular
distribution is normalized in such a way that 
\beq
\frac{d \Gamma}{dt} =
\int_{-1}^1 d(\cos \theta) \frac{d^2 \Gamma}{d \cos \theta~dt} = p(t)+m(t)~~~.
\eeq
As $t$ increases, one should see a growth of the $\sin^2 \theta$ component.

The derivation of Eq.~(\ref{eqn:dist}) is elementary.  The $\phi$ is coupled to
$K^+ K^-$ through an amplitude $\epsilon_\phi \cdot (p_{K^+} - p_{K^-})$, where
the quantities denote 4-vectors.  Thus the plane of (linear) $\phi$
polarization is related to that of the $K^+ K^-$ system in the $J/\psi$ rest
frame.  By definition, we have taken the $\phi$ linear polarization vector to
lie in the $x-y$ plane. 

One can decompose the decay amplitude $A$ into three independent components
\cite{FMJR}, corresponding to linear polarization states of the vector mesons
which are either longitudinal (0), or transverse to their directions of motion
and parallel ($\parallel$) or perpendicular ($\perp$) to one another. The
states $0$ and $\parallel$ are P-even, while the state $\perp$ is P-odd. Since
$J/\psi$ and $\phi$ are both C-odd eigenstates, the properties under P are the
same as those under CP. 

The case of transverse ($\parallel$ or $\perp$) polarization states is
reminiscent of photon polarization correlations \cite{Yang} in neutral pion
decay. Thus, an amplitude $A_\parallel$ (related to an interaction Lagrangian
proportional to ${\mathbf E}^2 - {\mathbf B}^2$) would have signified that the
$\pi^0$ had even CP, whereas the observed decay, involving the amplitude
$A_\perp$ and an interaction Lagrangian $\sim {\mathbf E} \cdot {\mathbf B}$,
signifies odd CP.  For the present case of massive vector mesons, there are
also longitudinal polarization states, corresponding to another even-CP
amplitude.

Consider the spatial components of the polarization three-vectors
$\bep_{J/\psi}$ and $\bep_\phi$ in the $J/\psi$ rest frame.  They must be
correlated since the decaying strange $B$ is spinless. The $J/\psi$ then has a
single linear polarization state $\bep$ for each amplitude. For longitudinally
polarized $\phi$ and $J/\psi$, characterized by the CP-even amplitude $A_0$, we
have $\bep_{J/\psi} = \hat x$. For transversely polarized $\phi$, with
$\bep_\phi = \hat y$, we have two possibilities: The CP-even amplitude
$A_\parallel$ corresponds to $\bep_{J/\psi} = \hat y$, while the CP-odd
amplitude $A_\perp$ corresponds to $\bep_{J/\psi} = \hat z$. 

A unit vector $n$ in the direction of the $\ell^+$ in $J/\psi$ decay may be
defined to have components 
\beq \label{eqn:ndef}
(n_x, n_y, n_z) = (\sin \theta \cos \varphi,
\sin \theta \sin \varphi, \cos \theta)~~~
\eeq
where $\varphi$ is the angle between the projection of the $\ell^+$ on the $K^+
K^-$ plane in the $J/\psi$ rest frame and the $x$ axis.  The sum over lepton
polarizations then leads to a tensor in the $J/\psi$ rest frame with spatial
components (in the limit of zero lepton mass, assumed here) 
\beq \label{eqn:lijdef}
\sum_{\ell^\pm {pol}} [\bar u \gamma_i v]^* [\bar u\gamma_j v]
\sim L_{ij} \equiv \delta_{ij} - n_i n_j~~~.
\eeq
Physically this tensor expresses the electromagnetic coupling of massless
lepton pairs to transverse polarization states of the $J/\psi$. 

When contracted with amplitudes corresponding to $|A_0|^2$, $|A_\parallel|^2$,
or $|A_\perp|^2$, the lepton tensor then yields terms $1 - n_x^2$, $1 - n_y^2$,
or $1 - n_z^2$, respectively.  The first two, when averaged over $\varphi$,
each give terms $(1 + \cos^2 \theta)/2$, while the last gives $\sin^2 \theta$.
We thus identify
\beq
p(t) = |A_0|^2 + |A_\parallel|^2~~~,~~~m(t) = |A_\perp|^2~~~.
\eeq

In the limit of flavor SU(3) symmetry, one expects the ratios of the relative
components in $B^0 \to J/\psi K^{*0}$ to be the same as those {\it at proper
time $t=0$} in the decays $B_s \to J/\psi \phi$ \cite{Isi93}. Thus, an analysis
of $B^0 \to J/\psi K^{*0}$ can provide an independent estimate of the relative
contributions of CP-even and CP-odd final states at $t=0$ to the decays $B_s
\to J/\psi \phi$, enhancing the ability to determine $\Gamma_H$ and $\Gamma_L$.
The dominance of the $|A_0|^2$ contribution in $B^0 \to J/\psi K^{*0}$ decays
\cite{CDFdist,ARGUS,CLEO} implies via flavor SU(3) that the $|A_0|^2$
contribution should also dominate $B_s \to J/\psi \phi$, and hence that
$B_s^{(-)} \to J/\psi \phi$ is likely to be suppressed in comparison with
$B_s^{(+)} \to J/\psi \phi$.  Thus the initial angular distribution is very
likely to be dominated by the $1 + \cos^2 \theta$ component. As time increases,
the fraction of the angular distribution proportional to this component will
decrease while that proportional to $\sin^2 \theta$ will increase.  It should
be possible to separate out the two components by a combined analysis in
$\theta$ and proper decay time.  If the $\sin^2 \theta$ component does not show
up even at large times, a single-exponential fit to the decay should provide a
good estimate of the lifetime of the CP-even eigenstate. 

The analysis performed by CDF \cite{CDFdist} for $B_s \to J/\psi \phi$
separated out the transverse component from the longitudinal component.  In the
absence of vertex cuts these would be, respectively, $\Gamma_T \equiv
\int_0^\infty dt (|A_\parallel|^2 + |A_\perp|^2)$ and $\Gamma_0 \equiv
\int_0^\infty dt (|A_0|^2)$.  With a minimum vertex cut of 50 $\mu$m, the
result obtained was $\Gamma_0/(\Gamma_0 + \Gamma_T) = 0.56 \pm 0.21
\mathrm{(stat)}_{~-0.04}^{~+0.02}\mathrm{(sys)}$. The transverse component
contains both CP-even and CP-odd contributions, while the longitudinal
component is CP-even. 

Corresponding determinations of $\Gamma_0/(\Gamma_0 + \Gamma_T)$ for the
decay $B^0 \to J/\psi K^{*0}$ are
$0.65 \pm 0.10 \pm 0.04$ (CDF) \cite{CDFdist},
$0.97 \pm 0.16 \pm 0.15$ (ARGUS) \cite{ARGUS},
$0.80 \pm 0.08 \pm 0.05$ (CLEO) \cite{CLEO}, and
$0.74 \pm 0.07$ (world average) \cite{CDFdist}.  This last value is compatible
with the corresponding one for $B_s \to J/\psi \phi$. A discrepancy would have
indicated either a violation of SU(3) or the lifetime effect mentioned above. 

\section{Enhancement of the $\Lambda_b$ decay rate}

The differences among lifetimes of particles containing heavy quarks are
expected to become smaller as the heavy quark mass increases and free-quark
estimates become more reliable.  Thus, mesons and baryons containing $b$
quarks are expected to have lifetimes differing no more than a few percent
\cite{NSNGU,Bigi}. For example, the process, $b u \to c d$ in the $\lb$ (``weak
scattering''), when considered in conjunction with the partially offsetting
process $b d \to c \bar u d d$ (``Pauli interference'') should lead to a small
enhancement in the $\lb$ decay rate, so that $\tlb = (0.9~\mathrm{to}~0.95)
\tb$.

The observed $\lb$ lifetime is $\tlb = 1.20 \pm 0.07$ ps, while the $B^0$ 
decays more slowly:  $\tb = (1.58 \pm 0.05)$ ps.  Here we have averaged a
compilation of world data \cite{Sharma} (for which $\tlb = 1.18 \pm 0.07$ ps)
with a new value \cite{CDFtb} $\tlb = 1.33 \pm 0.16 \pm 0.07$ ps.  The ratio of
these two quantities is $\tlb/\tb = 0.76 \pm 0.05$, indicating an enhancement
of the $\lb$ decay rate beyond the magnitude of usual estimates. 

We find \cite{JRLB} that, in spite of a large wave function for the
$b u$ pair in the initial baryon, which we denote by $\psbu$, only $(13 \pm
7)\%$ of the needed enhancement of the $\lb$ decay rate can be explained in
terms of the effects of the four-fermion matrix element. If we assume wave
functions are similar in all baryons with a single $b$ quark and two nonstrange
quarks, this quantity can be related to the hyperfine splitting $M(\sbs) -
M(\sb)$, for which the DELPHI Collaboration at LEP \cite{DELPHI} has recently
quoted a large value of $56 \pm 16$ MeV.  We estimate the effect of gluon
exchange by performing a similar calculation for $B$ mesons, relating the $B^*
- B$ splitting to the $B$ meson decay constant and taking account of differing
spin and hyperfine factors in the meson and baryon systems. 

A relation for the enhancement of the $\lc$ decay rate due to the weak
scattering process $c d \to s u$ was first pointed out in Ref.~\cite{BLS}.  At
the same order in heavy quark mass, one must also take account of Pauli
interference (interference between identical quarks in the final state)
\cite{VSGub}. Thus, for the $\lb$, one considers not only the process $b u \to
c d$ (involving matrix elements between $\lb$ states of $(\bar b b)(\bar u u)$
operators), but also those processes involving matrix elements of $(\bar b
b)(\bar d d)$ operators) which contribute to interference.  The net result of
four-quark operators in the $\lb$ is an enhancement of the decay rate by an
amount (see, e.g., Ref.~\cite{VSGub})
\beq \label{eqn:dg}
\dg(\lb) = (G_F^2/2 \pi) \psbu |V_{ud}|^2
|V_{cb}|^2 m_b^2(1-x)^2 [c_-^2 - (1+x)c_+(c_- - c_+/2)]~~~.
\eeq
We have neglected light-quark masses; $x \equiv m_c^2/m_b^2$, while $c_-$
and $c_+ = (c_-)^{-1/2}$ are the short-distance QCD enhancement and suppression
factors for quarks in a color antitriplet and sextet, respectively
\cite{QCDenh}: $c_- = [\alpha_s(m_c^2)/\alpha_s(M_W^2)]^\gamma$, where $\gamma
\equiv 12/(33 - 2 n_F)$, with $n_F = 5$ the number of active quark flavors
between $m_b$ and $M_W$. The $c_-^2$ term reflects the weak scattering process
$bu \to cd \to bu$; the remaining terms arise from destructive interference
between the two intermediate $d$ quarks in the process $bd \to c \bar u dd \to
bd$. 

Taking \cite{JRLB} $\alpha_s(m_b^2) = 0.193$ and $\alpha_s(M_W^2) = 0.114$, we
find $c_- = 1.32,~c_+ = 0.87$.  An estimate of $\psbu$ is then needed.  We find
it by comparing hyperfine splittings in mesons and baryons, under the
assumption that the strength of the one-gluon exchange term is the same for the
light quark -- heavy quark pair in each system.
Our result is that
\beq \label{eqn:ps}
\psbu = 2 \cdot \frac{2}{3} \cdot \frac{M(\sbs) - M(\sb)}{M(B^*) - M(B)}
\cdot \frac{M_B f_B^2}{12}~~~,
\eeq
where the first factor relates to color, the second to spin, and the last term
is the nonrelativistic estimate of the $b \bar u$ wave function in the $B$
meson \cite{wf}. (Here one may use the spin-averaged value of vector and
pseudoscalar masses for $M_M$.)  With the DELPHI value of $M(\sbs) - M(\sb)$,
the $B^* - B$ splitting of 46 MeV \cite{PDG}, and the estimate \cite{JRCP,lat}
$f_B = 190 \pm 40$ MeV, we obtain $\psbu = (2.6 \pm 1.3) \times 10^{-2}$
GeV$^3$. This is to be compared with $\psbub = M_B f_B^2/12 = (1.6 \pm 0.7)
\times 10^{-2}$ GeV$^3$ for the $B$ meson.

In both $\sb$ and $\sbs$, the light quarks are coupled up to spin 1. The
splitting then depends purely on the light quark -- heavy quark interaction.
The wave function between a light quark and a heavy one is assumed to be
identical in the $\lb$ and in the $\sb - \sbs$ system. The value of $\langle
\hat{s}_Q \cdot \hat{s}_{\bar q} \rangle$ is $(1/4,-3/4)$ for a
$(^3S_1,~^1S_0)$ $Q \bar q$ meson, where $Q$ and $q$ are the heavy and light
quark.  For a baryon $Qqq$ with $S_{qq} = 1$, one has $\langle \hat{s}_Q \cdot
\hat{s}_q \rangle = (1/4,~-1/2)$ for states with total spin (3/2,~1/2). Thus
the difference in $\hat{s}_i \cdot \hat{s}_j$ for the $\sbs - \sb$ splitting
(counting a factor of 2 for the two light quarks in the baryons) is 3/2 that
for the $B^* - B$ splitting. The factor of 2/3 in Eq.~(\ref{eqn:ps})
compensates for this ratio.  The color factor takes account of the fact that in
a meson, the quark and antiquark are a color ${\mathbf 3}$ and ${\mathbf 3}^*$
coupled to a singlet, while in a baryon the two ${\mathbf 3}$'s are coupled to
a ${\mathbf 3}^*$. 

In the relation (\ref{eqn:dg}) we now neglect $\sin \theta_c$ (setting $V_{ud}
=1$), and choose $m_b = 5.1$ GeV, $m_c = 1.7$ GeV, and $|V_{cb}| = 0.040 \pm
0.003$. We then find 
\beq
\dg(\lb) = 0.025 \pm 0.013~{ps}^{-1}~~~.
\eeq

The decay rates of the $B^0$ and $\lb$ are $\gb = 0.63 \pm 0.02$ ps$^{-1}$ and
$\glb = 0.83 \pm 0.05$ ps$^{-1}$, differing by $\dg(\lb) = 0.20 \pm 0.05$
ps$^{-1}$. The four-quark processes noted above can explain only $(13 \pm 7)\%$
of this difference, leading to an enhancement of only $(4 \pm 2)\%$ of the
total $\lb$ decay rate in contrast with the needed enhancement of $(32 \pm
8)\%$.

The corresponding calculation for charmed particles makes use of the
following inputs.

{\it 1. The $D$ meson decay constant} was taken \cite{JRCP} to be $f_D =
240 \pm 40$ MeV, leading to $\pscdb = (0.95 \pm 0.32)
\times 10^{-2}$ GeV$^3$;
{\it 2. The $D^* - D$ splitting} is assumed to be 141 MeV (the average for
charged and neutral states \cite{PDG});
{\it 3. Charmed baryon masses} are taken to be $M(\sc) = 2453$ MeV \cite{PDG}
and $M(\scs) = 2530 \pm 7$ MeV \cite{Ammosov};
{\it 4. The strong fine-structure-constant} at $m_c^2$ is taken to be
$\alpha_s(m_c^2) = 0.289$, consistent with the QCD scale mentioned above,
leading to $c_- = 1.60,~c_+ = 0.79$;
{\it 5. The strange quark mass} is taken to have a typical constituent-quark
value, $m_s = 0.5$ GeV.  We continue to neglect $u$ and $d$-quark masses for
simplicity;
{\it 6. The CKM factors} in Eq.~(\ref{eqn:dg}) undergo the replacements
$|V_{ud}|^2 |V_{cb}|^2 \to |V_{cs}|^2 |V_{ud}|^2$, which we approximate by
1 (again neglecting $\sin \theta_c$).

Our results for systems with $c$ and $b$ quarks are summarized in Table 1.
Remarks:

\begin{table}
\caption{Comparison of predicted squares of wave functions and decay rate
enhancements for $\lc$ and $\lb$.}
\begin{center}
\begin{tabular}{c c c} \hline
Quantity (units) & Charm & Beauty \\ \hline
$f_M$ (MeV)   &  $240 \pm 40$   &   $190 \pm 40$ \\
$|\Psi(0)|^2_{Q \bar q}~(10^{-2}$ GeV$^3$) & $0.95 \pm 0.32$
  & $1.6 \pm 0.7$ \\
$M(^3S_1) - M(^1S_0)$ (MeV) & 141 & 46 \\
$M(\Sigma^*) - M(\Sigma)$ (MeV) & $77 \pm 7$ & $56 \pm 16$ \\
$|\Psi(0)|^2_{Q q}~(10^{-2}$ GeV$^3$) & $0.69 \pm 0.24$
  & $2.6 \pm 1.3$ \\
$c_-$ & 1.60 & 1.32 \\
$c_+$ & 0.79 & 0.87 \\
$c_-^2 - (1+x)c_+(c_- - c_+^2/2)$ & 1.52 & 0.88 \\
$\dg(\Lambda_Q)$ (ps$^{-1}$) & $0.8\pm 0.3$ & $0.025 \pm 0.013$ \\ \hline
\end{tabular}
\end{center}
\end{table}

(a) The difference between the central values of $|\Psi(0)|^2_{Q \bar q}$ for
charm and beauty reflects the likely importance of $1/m_Q$ corrections (see,
e.g., Ref.~\cite{lat}), or -- in the language of the quark model -- of reduced
mass effects. 

(b) The $\scs - \sc$ hyperfine splitting used in this calculation is based on
one claim for observation of the $\scs$ \cite{Ammosov}, which requires
confirmation. 

(c) The value of $\psbu$ is large in comparison with the others for
light-heavy systems.  It would be helpful to verify the DELPHI $\sbs-\sb$
hyperfine splitting \cite{DELPHI}.  The ratio of hyperfine splittings for
charmed and beauty mesons is approximately 3:1, as expected if these splittings
scale as $1/m_Q$.  In contrast, the corresponding ratio for baryons is
considerably smaller, indicating a violation of $1/m_Q$ scaling. 

(d) The enhancement of the $\lc$ decay rate is quite modest. With $\glc \approx
5$ ps$^{-1}$, to be compared with $\gdn \approx 2.4$ ps$^{-1}$ and $\gdc
\approx 1$ ps$^{-1}$, one seeks an enhancement of at least $\glc - \gdn \approx
2.6$ ps$^{-1}$. If the enhancements $\dg(\Lambda_Q)$ in Table 1 were about a
factor of 4 larger, we could accommodate both the $\lc$ and $\lb$ decay rates,
but this is not consistent with our estimates of the matrix elements and their
effects on decay rates.  In particular, the effect of Pauli interference is to
cut the na\"{\i}ve estimate of the enhancement due to weak scattering alone
\cite{BLS} by roughly a factor of 2.

\section{Summary}

A combined analysis with respect to proper decay time and a single transversity
angle in the decay $B_s \to J/\psi \phi$ can determine the lifetime of at least
the CP-even and possibly the CP-odd mass eigenstates of the $B_s - \bar B_s$
system.  Additional information about the properties of the $J/\psi \phi$ mode
at proper time $t = 0$ can be obtained by a similar analysis of the decays $B^0
\to J/\psi K^{*0}$. Such analyses are in progress with the data samples
\cite{CDFdist,CLEO} now in hand. 

The DELPHI value \cite{DELPHI} of the $\sbs-\sb$ hyperfine splitting permits an
estimate of the overlap of quark wave functions between the $b$ quark and the 
light quarks in the $\Lambda_b$, and hence of the effect of four-quark
operators on its decay rate.  Even though the matrix element deduced from the
DELPHI result is quite large on the scale of those for heavy-light systems, one
can only account for $(13 \pm 7)\%$ of the difference between the $\lb$ and
$B^0$ decay rates. A similar approach also falls short of accounting for the
corresponding enhancement for the $\lc$ decay rate.

\section*{Acknowledgments}

I thank the Fermilab, CERN, and DESY Theory Groups and the Physics
Department of The Technion for their hospitality during various parts of this
work; the organizers of this workshop for partial support; Amol S. Dighe, Isard
Dunietz, and Harry J. Lipkin for an enjoyable collaboration on angular
distributions in decays; and B. Kayser, R. Kutschke, G. Martinelli, M. Neubert,
K. Ohl, M. Paulini, E. A. Paschos, C. T. Sachrajda, M. P. Schmidt, M. Shifman,
N. G. Uraltsev, and W. Wester for fruitful discussions. This work was supported
in part by the United States Department of Energy under Contract No.~DE FG02
90ER40560 and by the United States -- Israel Binational Science Foundation
under Research Grant Agreement 94-00253/1. 
 
% Journal and other miscellaneous abbreviations for references
% Phys. Lett. B style
\def \ajp#1#2#3{Am.~J.~Phys.~{\bf#1} (#3) #2}
\def \apny#1#2#3{Ann.~Phys.~(N.Y.) {\bf#1} (#3) #2}
\def \app#1#2#3{Acta Phys.~Polonica {\bf#1} (#3) #2 }
\def \arnps#1#2#3{Ann.~Rev.~Nucl.~Part.~Sci.~{\bf#1} (#3) #2}
\def \cmp#1#2#3{Commun.~Math.~Phys.~{\bf#1} (#3) #2}
\def \cmts#1#2#3{Comments on Nucl.~Part.~Phys.~{\bf#1} (#3) #2}
\def \cn{Collaboration}
\def \corn93{{\it Lepton and Photon Interactions:  XVI International Symposium,
Ithaca, NY August 1993}, AIP Conference Proceedings No.~302, ed.~by P. Drell
and D. Rubin (AIP, New York, 1994)}
\def \cp89{{\it CP Violation,} edited by C. Jarlskog (World Scientific,
Singapore, 1989)}
\def \dpff{{\it The Fermilab Meeting -- DPF 92} (7th Meeting of the American
Physical Society Division of Particles and Fields), 10--14 November 1992,
ed. by C. H. Albright \ite~(World Scientific, Singapore, 1993)}
\def \dpf94{DPF 94 Meeting, Albuquerque, NM, Aug.~2--6, 1994}
\def \efi{Enrico Fermi Institute Report No. EFI}
\def \el#1#2#3{Europhys.~Lett.~{\bf#1} (#3) #2}
\def \f79{{\it Proceedings of the 1979 International Symposium on Lepton and
Photon Interactions at High Energies,} Fermilab, August 23-29, 1979, ed.~by
T. B. W. Kirk and H. D. I. Abarbanel (Fermi National Accelerator Laboratory,
Batavia, IL, 1979}
\def \hb87{{\it Proceeding of the 1987 International Symposium on Lepton and
Photon Interactions at High Energies,} Hamburg, 1987, ed.~by W. Bartel
and R. R\"uckl (Nucl. Phys. B, Proc. Suppl., vol. 3) (North-Holland,
Amsterdam, 1988)}
\def \ib{{\it ibid.}~}
\def \ibj#1#2#3{~{\bf#1} (#3) #2}
\def \ichep72{{\it Proceedings of the XVI International Conference on High
Energy Physics}, Chicago and Batavia, Illinois, Sept. 6--13, 1972,
edited by J. D. Jackson, A. Roberts, and R. Donaldson (Fermilab, Batavia,
IL, 1972)}
\def \ijmpa#1#2#3{Int.~J.~Mod.~Phys.~A {\bf#1} (#3) #2}
\def \ite{{\it et al.}}
\def \jmp#1#2#3{J.~Math.~Phys.~{\bf#1} (#3) #2}
\def \jpg#1#2#3{J.~Phys.~G {\bf#1} (#3) #2}
\def \lkl87{{\it Selected Topics in Electroweak Interactions} (Proceedings of
the Second Lake Louise Institute on New Frontiers in Particle Physics, 15--21
February, 1987), edited by J. M. Cameron \ite~(World Scientific, Singapore,
1987)}
\def \ky85{{\it Proceedings of the International Symposium on Lepton and
Photon Interactions at High Energy,} Kyoto, Aug.~19-24, 1985, edited by M.
Konuma and K. Takahashi (Kyoto Univ., Kyoto, 1985)}
\def \mpla#1#2#3{Mod.~Phys.~Lett.~A {\bf#1} (#3) #2}
\def \nc#1#2#3{Nuovo Cim.~{\bf#1} (#3) #2}
\def \np#1#2#3{Nucl.~Phys.~{\bf#1} (#3) #2}
\def \pisma#1#2#3#4{Pis'ma Zh.~Eksp.~Teor.~Fiz.~{\bf#1} (#3) #2 [JETP Lett.
{\bf#1} (#3) #4]}
\def \pl#1#2#3{Phys.~Lett.~{\bf#1} (#3) #2}
\def \plb#1#2#3{Phys.~Lett.~B {\bf#1} (#3) #2}
\def \pr#1#2#3{Phys.~Rev.~{\bf#1} (#3) #2}
\def \pra#1#2#3{Phys.~Rev.~A {\bf#1} (#3) #2}
\def \prd#1#2#3{Phys.~Rev.~D {\bf#1} (#3) #2}
\def \prl#1#2#3{Phys.~Rev.~Lett.~{\bf#1} (#3) #2}
\def \prp#1#2#3{Phys.~Rep.~{\bf#1} (#3) #2}
\def \ptp#1#2#3{Prog.~Theor.~Phys.~{\bf#1} (#3) #2}
\def \rmp#1#2#3{Rev.~Mod.~Phys.~{\bf#1} (#3) #2}
\def \rp#1{~~~~~\ldots\ldots{\rm rp~}{#1}~~~~~}
\def \si90{25th International Conference on High Energy Physics, Singapore,
Aug. 2-8, 1990}
\def \slc87{{\it Proceedings of the Salt Lake City Meeting} (Division of
Particles and Fields, American Physical Society, Salt Lake City, Utah, 1987),
ed.~by C. DeTar and J. S. Ball (World Scientific, Singapore, 1987)}
\def \slac89{{\it Proceedings of the XIVth International Symposium on
Lepton and Photon Interactions,} Stanford, California, 1989, edited by M.
Riordan (World Scientific, Singapore, 1990)}
\def \smass82{{\it Proceedings of the 1982 DPF Summer Study on Elementary
Particle Physics and Future Facilities}, Snowmass, Colorado, edited by R.
Donaldson, R. Gustafson, and F. Paige (World Scientific, Singapore, 1982)}
\def \smass90{{\it Research Directions for the Decade} (Proceedings of the
1990 Summer Study on High Energy Physics, June 25 -- July 13, Snowmass,
Colorado), edited by E. L. Berger (World Scientific, Singapore, 1992)}
\def \smassb{{\it Proceedings of the Workshop in $B$ Physics at Hadron
Colliders} (Snowmass, CO, June 21 -- July 2, 1993), edited by P. McBride
and C. S. Mishra, Fermilab report Fermilab-CONF-93/267 (1993)}
\def \stone{{\it B Decays}, edited by S. Stone (World Scientific, Singapore,
1994)}
\def \tasi90{{\it Testing the Standard Model} (Proceedings of the 1990
Theoretical Advanced Study Institute in Elementary Particle Physics, Boulder,
Colorado, 3--27 June, 1990), edited by M. Cveti\v{c} and P. Langacker
(World Scientific, Singapore, 1991)}
\def \yaf#1#2#3#4{Yad.~Fiz.~{\bf#1} (#3) #2 [Sov.~J.~Nucl.~Phys.~{\bf #1} (#3)
#4]}
\def \zhetf#1#2#3#4#5#6{Zh.~Eksp.~Teor.~Fiz.~{\bf #1} (#3) #2 [Sov.~Phys. -
JETP {\bf #4} (#6) #5]}
\def \zpc#1#2#3{Zeit.~Phys.~C {\bf#1} (#3) #2}

\end{document}